\documentstyle[12pt]{article}
\title{FRAMES OF REFERENCE AND SOME OF ITS APPLICATIONS}
\author{Ll. Bel \\
Lab. Gravitation et Cosmologie Relativistes. ESA 7065 \\
{\small Tour 22-12, 4 place Jussieu, 75252 Paris}
}
\date{}
\begin{document}
\maketitle

\begin{abstract}

We define a Frame of reference as a two ingredients concept: A
meta-rigid motion, which is a generalization of a Born motion, and a
chorodesic synchronization, which is an adapted foliation. At the end
of the line we uncover a low-level 3-dimensional geometry
with constant curvature and a corresponding coordinated proper-time
scale. We discuss all these aspects both from the geometrical point
of view as from the point of view of some of the physical applications 
derived from them.

\end{abstract}

\section{Introduction}

Choosing a system of units is the first decision that a physicist has
to take to communicate the outcome of an experiment or the quantized
prediction of a theory. Metrology is therefore in some sense the more
universal branch of physics. 

Next in importance is to remind the frame of reference which has been
used to set up the experiment if it has already been done or to
describe the conditions under which the experiment has to be set. 

The systems of units have been chosen with the requirement that they
can be conveniently reproduced, or identified, to the desired
accuracy by anybody who has at his disposal the necessary equipment.
A family of allowed frames of reference has to be chosen keeping in
mind the same requirement but many other specific ones are necessary
to cope with the increased complexity of the concept. And more
important, not only is necessary to describe the hardware and the
protocols involved but also the theory which is supposed to describe
the concept.

Switching from a system of units to another is a very easy task.
Comparing data referred to different frames of reference may be a
very difficult one, depending on the domain which is being considered
and the theory which is available.

The concepts of {\it Absolute time} and {\it Absolute space} very
much simplify the theory of frames of reference in classical physics.
This theory relies heavily on the concept of {\it Rigid motions}, and
the behavior of solids as as objects which are able to move rigidly,
or approximately so at a desired level of accuracy,
under appropriate conditions.

Special and General relativity do not have a theory of Frames of
reference as complete and satisfactory as Classical physics. In
Special relativity only the theory of Galilean frames of reference or
that of irrotational Born motions is
firmly established. In General relativity only those space-times
possessing symmetries or Born congruences can be said to have frames
of reference everybody will agree upon. But in any case the number of
such particular frames of reference is never comparable to the number
offered by Classical physics.

There are two end of the line ingredients in the concept of a frame of
reference:  i) A time scale and ii) A low-level of geometry, i.e. the
geometry of the metrology.

As Helmholtz \cite{Helmholtz}, Poincar\'e \cite{Poincare} and Cartan 
\cite{Cartan} stressed,
this geometry has to possess the property of free mobility, meaning that
those objects that will be used as standards for measuring must
remain invariant when they move around. This is a very restrictive
property which requires this low-level geometry to be a Riemannian
geometry with constant curvature.

Of course the word geometry has different meanings depending on the
context on which it is used. The fact that in Relativistic physics one
has to do more often with space geometries which do not have constant
Riemannian curvature is by no means a contradiction. These geometries
are an essential part of the mathematical formalism and they should
be interpreted appropriately, but not necessarily as low-level
geometries. On the other hand the low-level geometry may be deeply
hidden in the formalism and has to be uncovered. 

Going from Absolute time of classical mechanics to a relativistic
time-scale based on proper-time coordination is a very innovative
aspect of Relativity physics but does not present any real conceptual
or mathematical difficulty. This is not so when abandoning the
concept of Absolute space the concept of the relativity of space is
considered with its implications in the process of measuring lengths,
surfaces or volumes. The culprit lies in the difficulty to define
rigid motions. The generalization of this concept is the first
problem that a consistent theory of the frames of reference has to
face.

This contribution is an update of early papers dealing with the
subject of frames of reference in Special and General relativity
written since 1990. It contains very little new material. The aim it
is more to look at some of the applications emphasizing what they
have in common and showing why a theory of frames of reference is
necessary to give a meaning to the problems these applications deal
with, or to produce numerical results everybody can understand.

The compilation that we present here includes an introduction to the
theory of Frames of reference in Relativity physics and the
presentation of five scaled down versions of the applications,
already published or posted, that this theory motivated. We hope that
understanding these applications will convince the reader of the
necessity of developing such a theory. 

Section 2 gives a rather detailed description of the theory of the
frames of reference introducing the concepts of quo-harmonic
congruences, meta-rigid motions, principal transformations, chorodesic
synchronizations and atomic time scale coordination, to which we
shall refer as {\it Temps Atomique Coordonn\'e (TAC)}. A previous
reading of this section is necessary to understand the vocabulary and
some of the fine points made in sections 2-7. But each of these
sections are independent one from the other.

Section 3 deals with the concept of strain from the unavoidable point
of view that requires that to define the strain of a body it is
necessary to compare it to another object, possibly an idealized one,
which serves as reference in the sense that it is harder to strain
than the object being tested. 

Section 4 is devoted to the analysis of the anisotropy of space in
a frame of reference co-moving with the Earth. It is an important
application because it leads to a prediction about the outcome of
experiments of the Michelson-Morley type when sensitivities of the
order of $10^{-12}-10^{-13}$ can be reached.

Section 5 introduces the concept of {\it Modes} of a quantized scalar
field in Milne's universe, \cite{Milne}. Although it is not more
difficult to deal with a general Robertson-Walker model we have
preferred to use this very simple model because besides the {\it TAC}
another important aspect plays a role in this case. Namely the
necessity of taking into account global properties of space-times
when selecting the family of allowed quo-harmonic congruences which
are acceptable as meta-rigid motions of frames of reference. 

Section 6 tackles the problem of identifying the mathematical meaning
of the family of transformations between two systems of quo-harmonic
coordinates adapted to two different frames of reference. We consider
as a simple example the family of Born congruences in Minkowski's
space-time.  

Section 7 is an essay at justifying a very simple cosmological model
which is based on an equation of state for a gas of zero mass
particles which differs from the usual one. Our equation of state is
deeply anchored in the theory of frames of reference of section 2. 

The Appendix is a short summary of definitions and notations
concerning the geometry of time-like congruences. The reader which is
not familiar with the subject is advised to start his reading with this 
appendix.

In all sections but the last (sect. 7) the system of units is
supposed to be such that the universal constant $c$ is equal to $1$.

\section{Frames of Reference}

We give below a precise and intrinsic definition of what we
mean by a {\it Frame of reference}, thus completing tentative definitions
presented in previous publications \cite{Bel90}, \cite{ABMM}, \cite{BMM},
\cite{Bel-Llosa1}, \cite{Bel-Llosa2}.

A frame of reference is a pair of geometrical objects
intrinsically defined in the space-time being considered: 
\begin{itemize}

\item A time-like congruence $\cal R$ of a particular type, that we
shall call the {\it Meta-rigid motion} of the frame of reference,
with its
corresponding {\it Principal transformation} which allows to
assign a constant number to each pair of world-lines, to be
interpreted as its physical distance, and to implement the free
mobility of ideal rigid bodies, thus uncovering a low-level geometry. 

\item A space-like foliation of a particular type $\cal F$, that we
shall call the {\it Chorodesic synchronization} of the frame of
reference, with its corresponding ({\it Atomic time scale
coordination}).

\end{itemize}

\subsection{\it Meta-rigid motions}. 

In classical physics rigid motions are defined as
those congruences which keep constant the euclidean distance between
any two simultaneous events with respect to absolute time. That is to
say:

\begin{equation}
\label {2.32}
t=t, \quad x^i(t)= R^i_j(t)\left[z^j- A^j(t)\right],
\end{equation}  
where $x^i(t)$ and $z^i$ are cartesian coordinates and
$R^i_j(t)$ are rotation matrices.
Two main properties of these congruences are worth mentioning here
besides some other obvious ones: 

i) Let $v^i(t,z^j)$ be the velocity field of any congruence.
Then rigid congruences can be characterized by the local condition:

\begin{equation}
\label {2.33}
\sigma_{ij}=\partial_i v_j+\partial_j v_i=0
\end{equation}  

ii) the knowledge of $a^j(t)=\dot v^j$, where the dot means a
derivative with respect to time, and the rotation rate 
$\omega_{ij}=\partial_i v_j-\partial_j v_i$ along one of the
world-lines defines the whole congruence.

iii) The rigid-motion congruences are homogeneous in the sense that
none of the world-lines is privileged in the definition, and every one
together with their corresponding rotation rate field may be
considered to be the seed of the same congruence.

To define the type of admissible meta-rigid motions $\cal R$ of
relativistic physics three general conditions have to be imposed
which are meant to make this concept to inherit as much as
possible of the formal properties of rigid motions in classical
physics: 

i) Their local characterization must be intrinsic and have a
meaning independently of the space-time being considered. But
boundary and global conditions must be appropriate to each particular case.

ii) The knowledge of any open sub-bundle of the congruence
must be sufficient to characterize the whole congruence.

iii) The congruence has to be homogeneous, i.e. its definition does not 
have to distinguish any particular world-line of the congruence.

To comply with these three conditions the most natural
approach is to require the vector field $u^\alpha$ of the
meta-rigid motions to satisfy a set of differential conditions.
Obvious candidates which satisfy this requirement are the differential
equations defining the Killing congruences.

\begin{equation}
\label {2.10}
\Sigma_{ij}=\hat\partial_t \hat g_{ij}=0, \quad
\partial_i \Lambda_j-\partial_j \Lambda_i =0.
\end{equation}

Other congruences to be acceptable as meta-rigid motions
are the Born congruences, which are a generalization of
the Killing congruences and are defined by the single group of
conditions:

\begin{equation}
\label {2.11}
\partial_t \hat g_{ij}=0.
\end{equation}

We shall continue to call these congruences Rigid congruences, as usual.
Meta-rigid ones will be an appropriate generalization of them. 

The Born conditions, and a fortiori the Killing conditions, are very
restrictive in any space-time including Minkowski's one,
\cite{Herglotz}, \cite{Noether}, and the question may be raised of
generalizing the concept of meta-rigid motion in such a way that the
family of congruences $\cal R$ generalize Born congruences up to the
point of having hopefully a number of degrees of freedom equivalent
to that of the group of rigid motions in classical
mechanics\,\footnote{In references \cite{Bel-Llosa1} and
\cite{Bel-Llosa2} some other apparently natural generalizations were
suggested along this line but they all looked too restrictive in
three dimensions of space}.

When facing this problem, usually forced by a physical application
which requires the sublimated modelisation of the behavior of a
rigid body, many authors recur to use Fermi congruences in its
general form, \cite{Mast}. By this we
mean the time-like congruences which world-lines have constant Fermi
coordinates
based on a distinguished world-line seed. Because of this
distinction, except in the case where they are in fact Born
congruences, Fermi congruences are not acceptable candidates to the
concept of a meta-rigid motion \cite{Salas}.   

Harmonic congruences can be defined as those time-like congruences
$u^\alpha$ for which there exist three space-like functions
$f^a(x^\alpha)$ satisfying the following equations:

\begin{equation} 
\label {2.12} 
\Box f^a=0, \quad
u^\alpha\partial_\alpha f^a=0, \quad a,b,c=1,2,3 
\end{equation} 
where $\Box$ is the d'Alembertian operator corresponding to
the space-time metric.

Harmonic congruences, with the corresponding harmonic coordinates,
have been used extensively in the literature for several reasons,
including mainly the fact that they simplify many calculations, even
if now and then some inconsistencies have been pointed out in some
particular problems \cite{Madore}. Harmonic congruences could in
principle be considered good representatives of meta-rigid motions.
In fact they are not. They are acceptable generalizations of Killing
congruences but they are not generalizations of the Born
congruences.  More precisely it has been proved \cite{Bel-Coll} that
irrotational Born congruences, which are not Killing, are never
Harmonic congruences. A familiar example of such congruence is the
Born congruence generated by an arbitrary non geodesic world-line of
Minkowski space-time.

To solve the problem just mentioned concerning the harmonic
congruences it has been proposed to identify the meta-rigid motions
with appropriately selected {\it Quo-harmonic congruences},
\cite{Salas}, \cite{Bel-Llosa1}, \cite{Bel-Llosa2}. The latter being
those congruences for which there exist three independent space-like
solutions $f^a(x^\alpha)$ satisfying the following equations:

\begin{equation}
\label {2.15}
\hat\triangle f^a =0
\end{equation} 
where:

\begin{equation}
\label {2.14}
\hat\triangle \equiv \hat g^{ij}(\hat\partial_i\hat\partial_j 
- \hat\Gamma^k_{ij}\hat\partial_k)
\end{equation}
and where $\hat\Gamma^k_{ij}$ are the symbols defined in \ref{2.8}.
These are proper generalizations of Born
congruences in the sense that any Born congruence is also a
quo-harmonic one because in this case $\hat g_{ij}(x^k)$ does not
depend on time and the operator $\hat \triangle$ is the usual
Laplacian operator corresponding to a 3-dimensional Riemannian
metric. It has been impossible up to now to determine, or even to
estimate, the number of degrees of freedom of the class of
quo-harmonic congruences in a general space-time except that it is
certainly more general than that of Born congruences and it even
looks too general, letting us to think that global conditions,
boundary conditions, or supplementary physical conditions, should be
invoked to restrict this generality.

Notice that if a congruence is quo-harmonic and quo-harmonic
coordinates are used, $x^i=f^i$, then:

\begin{equation}
\label {2.15bis}
\hat g^{ij}\hat\Gamma^k_{ij}=0
\end{equation} 
Obviously, if adapted space coordinates of a congruence can be found
such that the preceding conditions are satisfied then this congruence
is quo-harmonic.

\subsection{\it Principal transformations of the Fermat quo-tensor}

Let us consider the Zel'manov-Cattaneo tensor \ref{2.9}.  
We shall say that a skew-symmetric
quo-tensor $\hat f^{ij}$ is an eigen quo-form if it 
gives a stationary value to the function:

\begin{equation}
\label {2.17}
\sigma \equiv - {{\hat R_{ijkl}\hat f^{ij}\hat f^{kl}}  \over
{\hat g_{ijkl}\hat f^{ij}\hat f^{kl}}}
\end{equation}
where:

\begin{equation}
\label {2.18} 
\hat R_{ijkl} = \hat g_{is}\hat R^s_{jkl}, \quad
\hat g_{ijkl} = \hat g_{ik}\hat g_{jl}-\hat g_{jk}\hat g_{il}
\end{equation}
which will be by definition the corresponding eigen-value.
Equivalently we can consider this other tensor, \cite{Ida} and
\cite{Bel90}:  

\begin{equation}
\label {2.19}
\tilde R_{ijkl} = \frac14 (\hat R_{ijkl} - \hat R_{jikl} 
+ \hat R_{klij} - \hat R_{lkij})
\end{equation}
Since this quo-tensor has the symmetries of a Riemann tensor in three 
dimensions we have the following well known identity, which expresses that
the Weyl-like object is zero:

\begin{equation}
\label {2.20}
\tilde R_{ijkl}=\hat g_{ik}\tilde R_{jl} - \hat g_{il}\tilde R_{jk}
+ \hat g_{jl}\tilde R_{ik} - \hat g_{jk}\tilde R_{il}
-\frac12 R(\hat g_{ik}\hat g_{jl}-\hat g_{il}\hat g_{jk})
\end{equation}
where:

\begin{equation}
\label {2.21}
\tilde R_{ik} = \hat g^{jl}\tilde R_{ijkl}, \quad 
\tilde R = \hat g^{ik}\tilde R_{ik}
\end{equation}
A simple calculation shows then that $\sigma$ in \ref{2.17} is equal to:

\begin{equation}
\label {2.22}
\sigma= {{\tilde S_{ij} \hat n^i\hat n^j} \over {\hat g_{kl}\hat n^k\hat
n^l}}
\end{equation}
where:

\begin{equation}
\label {2.23}
\tilde S_{ij} \equiv \tilde R_{ij}-\frac12 \tilde R \hat g_{ij}, \quad 
\hat n^i=\delta^{ijk}_{123}\hat f_{jk}
\end{equation}
$\delta^{ijk}_{abc}$ being the Kronecker tensor of rank six.
Therefore the eigen-forms of the Zel'manov-Cattaneo tensors are the duals
of the eigen-vectors of the Einstein-like quo-tensor $\tilde S_{ij}$ and
both objects
have the same eigen values $\sigma_a$.

Let $\hat n^i_a$ be three orthonormal eigen-vectors of $\tilde S_{ij}$,
with
corresponding eigen-values $\sigma_a$:

\begin{equation}
\label {2.24}
\hat g_{ij}=\epsilon^a\hat n_{ai}\hat n_{aj}, \quad 
\tilde S_{ij}\hat n^j_a=\sigma_a\hat n_{ai}, \quad
\hat n_{ai}=\hat g_{ij}\hat n^j_a, \quad \epsilon^a=1. 
\end{equation}
We shall say that a 3-dimensional quo-tensor $\bar g_{ij}$ is a {\it
Principal transform}\,\footnote{This concept was defined in a more
restricted environment in \cite{Principal}.} of the Fermat quo-tensor
$\hat g_{ij}$ if there exist three functions $c_a(x^\alpha)$ such
that the following quo-tensor:

\begin{equation}
\label {2.34}
\bar g_{ij}=\epsilon^a \bar n_{ai}\bar n_{aj}, \quad  
\bar n_{ai}=c_a\hat n_{ai}
\end{equation}
has the following properties:

\begin{itemize}
\item It is independent of time:

\begin{equation}
\label {2.35}
\partial_t \bar g_{ij}=0
\end{equation}

\item The Riemannian metric that it defines has constant curvature:

\begin{equation}
\label {2.25}
\bar R_{ijkl}=\kappa(\bar g_{ik}\bar g_{jl}-\bar g_{il}\bar g_{jk})
\end{equation} 

\item If $\sigma_a = \sigma_b$ then $c_a=c_b$ 

\item Satisfies the quo-harmonic condition:

\begin{equation}
\label {2.27}
(\hat\Gamma^i_{jk}-\bar\Gamma^i_{jk})\hat g^{jk}=0
\end{equation}
where the $\bar\Gamma^i_{jk}$'s are the Christoffel connection symbols
corresponding
to the metric $\bar g_{ij}$. 
\end{itemize}

If $\kappa$ above is zero then the metric $\bar g_{ij}$ is euclidean
and the last condition \ref{2.27} implies that cartesian coordinates
of this metric, which are the most convenient adapted coordinates to
use, are a system of quo-harmonic coordinates of the Fermat
quo-tensor. If $\kappa \not=0$ then, even when the congruence is
quo-harmonic, quo-harmonic coordinates are not necessarily the most
convenient adapted coordinates to use. This is the case for instance
when the meta-rigid motion is the standard conformal Killing
congruence of reference of a Robertson-Walker model. 

To interpret the three frame of reference dependent scalars $c_a$ and
directions $\bar n_{ai}$ we proceed as follows. Let us consider a
light signal, or any other signal, which propagates along a null curve. Let
us
assume that it leaves a location on a world-line $W_1$ with adapted
coordinates $x^i$ at time $t$, and it reaches a location on a
world-line $W_2$ with adapted coordinates $x^i+dx^i$ at time
$t+dt$ where it bounces back intersecting again the world-line
$W_1$ at an event with coordinates $x^i$ and $t+2d\tau$. We have then:

\begin{equation}
\label {2.55}
g_{\alpha\beta}(x^\rho)dx^\alpha dx^\beta \quad \hbox{or} \quad
\epsilon_a (\theta^a)^2=(\theta^0)^2
\end{equation}
with:

\begin{equation}
\label {2.56}
\theta^0=u_\alpha dx^\alpha=d\tau, \quad \theta^a=\hat n^a_i dx^i,
\quad \hat n^a_i=\hat n_{ai} 
\end{equation}
where $d\tau$ is also:

\begin{equation}
\label {2.58}
d\tau=\sqrt{\hat g_{ij}dx^idx^j}
\end{equation}
The distance between $W_1$ and $W_2$ according to our interpretation
of the constant curvature metric $\bar g_{ij}$ is:

\begin{equation}
\label {2.57}
D=\sqrt{\bar g_{ij}dx^idx^j}
\end{equation}
Therefore the {\it round-trip speed} of the signal is:

\begin{equation}
\label {2.59}
v=\sqrt{\frac{\bar g_{ij}dx^idx^j}{\hat g_{ij}dx^idx^j}}
\end{equation}
If: 

\begin{equation}
\label {2.60}
dx^i=\bar k\beta^a \bar n_a^i, \quad \hbox{with} \quad 
\bar n_a^i=\bar g^{ij}\bar n_{aj}, \quad \delta_{ab}\beta^a\beta^b=1 
\end{equation} 
where $\bar k$ is any small factor, then we obtain:

\begin{equation}
\label {2.61}
v^{-1}=\sqrt{\frac{\beta_1^2}{c_1^2}+\frac{\beta_2^2}{c_2^2}+
\frac{\beta_3^2}{c_3^2}}, \quad \beta_a=\beta^a
\end{equation} 
If instead we represent the direction of propagation by:

\begin{equation}
\label {2.62}
dx^i=\hat k\gamma^a \hat n_a^i, \hbox{with} \quad 
\hat n_a^i=\hat g^{ij}\hat n_{aj}, \quad
\delta_{ab}\gamma^a\gamma^b=1 
\end{equation}
then we obtain the equivalent result:

\begin{equation}
\label {2.63}
v=\sqrt{\gamma_1^2 c_1^2+\gamma_2^2 c_2^2+\gamma_3^2 c_3^2},
\quad \gamma_a=\gamma^a
\end{equation}

It follows from \ref{2.61} or \ref{2.63} that the principal directions
$\bar n_a$ are the directions where the round-trip speed of a light signal
is stationary and that the stationary values are given by the
scalars $c_a$. 

\subsection{\it Chorodesic synchronizations}. The choice of a
foliation to define the synchronization of a frame of reference is to
some extent less essential than the choice of a quo-harmonic
congruence to define its meta-rigid motion.  Besides, time is a much
better understood concept in General relativity than the concept of
space. The role of a synchronization is to be a first step towards a
convenient definition of a universal scale of time, and the
properties that are to be required from a synchronization will depend
on the use for which it is intended. Atomic and sidereal time are two
scales of common use which correspond to different synchronizations of
the frame of reference co-moving with the Earth. Not to mention many
other scales of time used in astronomy. To illustrate this point we
consider briefly how the {\it International Atomic Time (TAI)} scale
is defined. It is based on the definition of the second in the
international system of units {\it SI} as a duration derived from the
frequency of a particular atomic transition.   

This definition is universal, in the sense that any physicist is able
to use a well defined second, but at the same time is local because
the identification of atomic time with proper-time duration implies
that this definition of the second can be only used to define a scale
of time along the world-line of the clock being used used as a
standard reference.

To define a scale of time on Earth to be used on navigation systems
like the GPS for instance requires a different approach. In this
case\,\footnote{See for instance \cite{Soffel}} the second is defined
as above with some precisions added. The resulting definition of {\it
TAI} is:

{\it TAI is a coordinate time scale defined in a geocentric reference
frame with an SI second realized on the rotating
geoid}.\,\footnote{The geoid is a level surface of constant
geo-potential (gravity and centrifugal potential) at ``mean sea
level", extended below the continents.}

It follows from this that the operational definition of
the second is relative to some particular locations, i.e. world-lines
of a particular congruence, which in this case is the rotating Killing
congruence corresponding to the frame of reference co-moving with the
Earth. A second at some other location is then defined as to nullify
the relativistic red-shift between any pair of clocks of reference
co-moving with the Earth. The second at any other location, not in
the geoid, is then the interval of time separating the arrival of
two light signals sent, one second apart, from a standard clock
located on the geoid. The red-shift formulas of General relativity
allow then to compare the rhythm of any two clocks that can be joined
with light signals.

Let $\cal R$ be any time-like congruence. A {\it Chorodesic} $C$ of $\cal
R$ is by definition, \cite{Salas} \cite{Bel-Llosa2}, a line such that its
tangent vector $p^\alpha$
satisfy the following equations:

\begin{equation}
\label{2.53}
\frac{dx^\alpha }{d\lambda }=p^\alpha, \quad 
\frac{\nabla p^\alpha 
}{d\lambda } =\frac 12u^\alpha \Sigma _{\mu \nu }p^\mu p^\nu , 
\end{equation}
where $\lambda$ is the proper length along $C$, and
$\Sigma_{\mu\nu}$ is the covariant expression of Born's
deformation rate \ref{8.12}. Obviously if $\Sigma _{\mu \nu }=0$, i.e.
if $\cal
R$ is a Born congruence then the chorodesics of $\cal R$ and the
geodesics of the space-time coincide.

Chorodesics of a congruence are important mainly because of the
following result:

{\it If a space-like chorodesic $C$ is orthogonal to a world-line of
a congruence $\cal R$ then it is orthogonal to all the world-lines of the
congruence $\cal R$ that it crosses.}

This follows from deriving the scalar product $p^\rho u_\rho$ along
$C$. We	thus get:

\begin{equation}
\label {2.50}
\frac{d}{d\lambda}(p^\rho u_\rho)=
-\frac 12\Sigma _{\mu \nu }p^\mu p^\nu+
\frac12 p^\rho p^\sigma(\nabla_\rho u_\sigma+\nabla_\sigma u_\rho)  
\end{equation}
and using the definition of $\Sigma _{\mu \nu }$ this is equivalent
to:

\begin{equation}
\label {2.51}
\frac{d}{d\lambda}(p^\rho u_\rho)=-2(p^\rho u_\rho)(p^\rho \Lambda_\rho)
\end{equation}
from where the result above follows.

Particular foliations associated with a congruence $\cal R$ are the
one parameter family of hyper-surfaces generated by the
chorodesics orthogonal to any particular world-line of the
congruence. We call these foliations {\it Chorodesic
synchronizations.} In general a chorodesic synchronization depends on
a world-line seed. But if the congruence is integrable then all its
chorodesic synchronizations coincide with the family of hyper-surfaces
orthogonal to the congruence. 

Let $\cal C$ be the chorodesic synchronization orthogonal to some
world-line $W$ of a frame of reference $\cal R$. Then the associated
{\it Atomic (or Proper) Time Coordination (TAC)} scale is by
definition a time coordinate which value at any event $E$ is the
proper time interval between some arbitrary event on $W$ and the
intersection with $W$ of the chorodesic hypersurface of $\cal C$
containing $E$.

Notice that, even when the congruence of reference $\cal R$ is
hypersurface orthogonal, i.e. there exists a single chorodesic
synchronization, the {\it TAC} may depend on the particular
world-line $W$ which is used to identify the coordinate time with
proper-time along $W$. Switching from a world-line seed of a {\it
TAC} to another gives rise to a sub-group of adapted coordinate
transformations that we shall call the {\it TAC sub-group}.  An
explicit example will be shown in section 6.  A complete definition
of the {\it TAC} requires therefore to choose a particular world-line
$\cal S$, or eventually a bunch of equivalent ones. On Earth the {\it
TAC}, i.e. the {\it TAI}, is associated to the bunch of world-lines
$\cal S$ of locations on the geoid.

We explain below in geometrical terms why this allows to define a
convenient {\it TAC}, in any stationary frame of reference, not
necessarily co-moving with the Earth. Let us assume that $\cal R$ is
a Killing congruence. In this particular case the {\it TAC} with seed
on any particular world-line of $\cal R$ implements on any other
world-line of $\cal R$ the local time scale correctly red-shifted. In
fact in this case the chorodesics of $\cal R$ are geodesics, and the
flow of geodesics is invariant under the group of motions generated
by the Killing vector field of the Killing congruence.  Let $A_1$ and
$B_1$ be two events on a world-line $W_1$ of $\cal R$ and let $A_2$
and $B_2$ be the events at which the geodesic hyper-surfaces
orthogonal to $W_1$ intersect a second world-line $W_2$ of the
congruence. Let $L_2$ and $M_2$ be the intersections of the future
(resp. past) light-cones with origins $A_1$ and $B_1$ with $W_2$. It
follows then very easily that the proper-time interval $A_2-B_2$ on
$W_2$ is the same as $L_2-M_2$. 

If $\cal R$ is not a Killing congruence then the preceding result
does not hold, nevertheless general {\it TAC} scales, as we defined
them, remain useful as theoretical time references. Its practical use
remains to be understood.

\section{Elastic deformations}

The main ingredients in the theory of elastic deformations are:
\begin{itemize}
\item 1.- A definition of the infinitesimal displacement vector.
\item 2.- The kinematical description of the strain tensor field as a
measure of the
deformation relative to some idealized rigid body of reference. 
\item 3.- The dynamics of the stress tensor and the equations of
equilibrium   
\item 4.- The Hooke's law
\item 5.- The Beltrami-Michell integrability conditions
\end{itemize}

In ordinary elasticity theory items 1 and 2 rely heavily on the
geometry of euclidean space. First of all in using the concept of
rigid motion as a motion of reference and second in measuring strains
in terms of change of euclidean distances between corresponding
points of the initial body and the deformed one. Some of the
definitions below make use of the concepts introduced in the
preceding section as substitutes to the elementary ones used in
ordinary elasticity theory.

Let us consider an elastic body which motion is described by a
congruence $\cal B$ with unit tangent vector-field $w^\alpha$.
Suppose that among the motions of reference, i.e. the meta-rigid
motions of the space-time we are considering, there is one $\cal R$,
such that i) One of the world-lines $W$ of $\cal R$ coincides with one of
the world-lines of $\cal B$ and the rotation rates of both
congruences along $W$ coincide, ii) there exists a one to one
correspondence between the world-lines of $\cal B$ and a subset of
world-lines of $\cal R$ such that the orthogonal distance, in the
sense of the principal transform metric $\bar g_{ij}$ of the Fermat tensor,
between corresponding world-lines, remains small. Then we shall say
that the motion of $\cal B$ is a deformation of $\cal R$ and define
this kinematic deformation in terms of the geometry of the meta-rigid
motion $\cal R$.

Let us consider a system of coordinates adapted to $\cal R$ and let $\cal
B$
be defined in this system of coordinates by the parametric equations:

\begin{equation}
\label {3.1}
y^\alpha = x^\alpha + \zeta^\alpha(x^\rho)
\end{equation}
$x^0$ being the parameter. A short and explicit calculation proves
that if $u^\alpha$ is the unit tangent vector to the congruence $\cal
R$ and we use the following notation:

\begin{equation}
\label {3.2}
w^\alpha = u^\alpha + \delta u^\alpha
\end{equation}
then one has in this system of adapted coordinates:

\begin{equation}
\label {3.3}
\delta u^0 = -\xi^{-1} \varphi_i \partial_0 \zeta^i, \quad
\delta u^i = \xi^{-1}\partial_0 \zeta^i,
\end{equation}
a result which can be written in the manifestly covariant form:

\begin{equation}
\label {3.4}
\delta u^\alpha = \hat g^\alpha_\lambda {\cal L}(\zeta)u^\lambda  
\end{equation}
where $\cal L()$ is the Lie derivative operator. It follows from this
result that the generator of a deformation is defined up to the
transformation:

\begin{equation}
\label {3.5}
\zeta^\alpha \rightarrow \zeta^\alpha + k u^\alpha
\end{equation}
where $k$ is any function. This property can thus be used to assume,
without
any loss of generality, that $u^\alpha$ and $\zeta^\alpha$ are orthogonal:
$u_\alpha \zeta^\alpha = 0$. Or in adapted coordinates:

\begin{equation}
\label {3.6}
\zeta_0=0
\end{equation} 

By definition the strain tensor of the congruence $\cal B$ with respect
to the frame of reference $\cal R$ is, using a system of coordinates
adapted to it, the tensor:

\begin{equation}
\label {3.7}
\epsilon_{ij}= \bar\nabla_i \zeta_j + \bar\nabla_j \zeta_i 
\end{equation}
where:

\begin{equation}
\label {3.7bis}
\bar\nabla_i \zeta_j=\hat\partial_i\zeta_j-\bar\Gamma^k_{ij}\zeta_k  
\end{equation}
This definition is strongly dependent on the theory of the frames of
reference presented in the preceding section.
Notice also that $\epsilon_{ij}$ defines $\zeta_i$ only up to a
transformation:

\begin{equation}
\label {3.8}
\zeta_\alpha \rightarrow \zeta_\alpha + s_\alpha
\end{equation} 
where $s_\alpha$ is a solution of the following equations:

\begin{equation}
\label{3.9}
\bar\nabla_i s_j + \bar\nabla_j s_i = 0
\end{equation}

If the deformation is a {\it static} one, this
meaning that $\delta u^\alpha=0$ in \ref{3.2}, then from \ref{3.3} 
it follows that:

\begin{equation}
\label{3.9bis}
\partial_0 \zeta^i=0
\end{equation}
If moreover $\bar g_{ij}=\hat g_{ij}$, i.e. if the Fermat tensor is a
Riemannian metric with constant curvature, then the definition of
strain in \ref{3.7} is the usual one of ordinary elasticity
theory, and Eqs. \ref{3.9} are the equations defining the
infinitesimal generators of ordinary rigid motions of classical
physics. This behavior in these particular cases is our
justification for the definition \ref{3.7}.

The stress-energy of an elastic test body is:

\begin{equation}
\label {3.10}
T^{\alpha\beta}=\rho w^\alpha w^\beta -\pi^{\alpha\beta}, \quad
\pi^\alpha_\beta w^\beta=0
\end{equation} 
where $\rho$ is its density, $w^\alpha$ is its unit four-velocity
field, and where $\pi_{\alpha\beta}$ is the tensor describing the
stresses which are present as a result of the volume forces acting on
the body and those acting on its surface. Its components have to be
considered to have the same order of smallness as that of $\delta
u^\alpha$ and therefore the second Eq. \ref{3.10} is equivalent to
$\pi_{\alpha\beta}u^\alpha=0$. The generalization of the
equilibrium equations of ordinary elasticity theory to a relativistic
one is then, assuming that no other forces that gravitational or
inertial ones act on the body:

\begin{equation}
\label {3.11}
\nabla_\alpha T^\alpha_\beta=0,
\end{equation}
This ingredient is common to any relativistic elasticity theory  and
owes nothing to any theory about the frames of reference.

The formulation of Hooke's law, which is the next ingredient needed
in the theory, depends again from a conceptual point of view on the
the choice of the metric used to compare distances between points of
a body and the same body after, or during, the action of some
additional stresses. Our present version of Hooke's law for
homogeneous and isotropic bodies is:

\begin{equation}
\label {3.12}
\epsilon_{ij}={1 \over {3(3\lambda+2\mu)}}\pi\bar g_{ij}+
{1\over{2\mu}}(\pi_{ij}-\frac13 \bar g_{ij}\pi), \quad 
\pi=\bar g^{ij}\pi_{ij}
\end{equation} 
where $\lambda$ and $\mu$ are the Lam\'e's parameters of the body.

The Beltrami-Michell equations, which is the last ingredient
we mentioned at the beginning of this section, are the equations that
should be imposed to the stress-tensor $\pi_{ij}$ to satisfy
the integrability conditions corresponding to the definition \ref{3.7}
of the strain tensor, the conservation equations \ref{3.11}, and Hooke's
law. They are cumbersome and they will not be written here.

In \cite{Laguna} the Fermat tensor instead of a principal
transformation of it was used in Eqs. \ref{3.7} and \ref{3.12}. The
choice which we have made here agrees better with our present
confidence on the meaning and the necessity of considering the
metric $\bar g_{ij}$ to implement the free mobility property of
idealized rigid
bodies.

\section{Local anisotropy of space in a frame of reference co-moving with 
the Earth} 

We consider in this section\,\footnote{This section has been included
here for completeness. It is an abstract of a longer contribution by
Ll. Bel and A. Molina to this same volume. A full version of this
work can be seen also in \cite{Bel-Molina}. We use natural units such
$G=c=1$} the linearized line-element
of the gravitational field of the earth in a co-moving
frame of reference, and we construct the euclidean principal
transformation of the corresponding Fermat tensor. This application
is particularly important because it leads to a prediction that can be
tested
with available technology.

The line-element of the linearized exterior gravitational field of a 
non-rotating spheroidal body with mass $M$, mean radius $R$, and reduced 
quadrupole moment $J_2$ is:

\begin{equation}
\label {4.1}
ds^2=-(1-2U_G)dt^2+(1+2U_G)(dr^2+r^2(d\theta^2
+\sin^2\theta d\varphi^2))
\end{equation}
where $U_G$ is:

\begin{equation}
\label {4.2}
U_G=\frac{M}{r}\left(1+\frac{1}{2}\frac{J_2R^2(1-3\cos^2 \theta
)}{r^2}\right)
\end{equation}

If the body is rotating with angular velocity $\Omega$ then we obtain
the line-element substituting in \ref{4.1} $\varphi+\Omega t$ for
$\varphi$.
The line element of the Earth at the approximation we are interested
in is therefore:

\begin{equation}
\label {4.2}
ds^2=-(1-2(U_G+U_\Omega))dt^2+
2 A_\varphi dt d\varphi
+(1+2U_G)(dr^2+r^2( d\theta^2
+\sin^2 \theta d\varphi^2))
\end{equation}
where:

\begin{equation}
\label {4.3}
U_\Omega=\frac12\Omega^2 r^2\sin^2 \theta , \quad 
A_\varphi=\Omega r^2\sin^2 \theta
\end{equation}
The Fermat line-element of space can then be written as:

\begin{equation}
\label {4.4}
d\hat s^2=(\hat\theta^1)^2+(\hat\theta^2)^2+(\hat\theta^3)^2
\end{equation}
where:

\begin{equation}
\label {4.5}
\hat\theta^1=(1+U_G)(dr+J_2 f rd\theta), \,\,
\hat\theta^2=(1+U_G)(-J_2 f dr+ rd\theta), \,\,
\hat\theta^3=(1+U)r\sin \theta d\phi
\end{equation}
with:

\begin{equation}
\label {4.6}
f=-4\frac{R^2}{r^2}\left(1-q\frac{r^5}{R^5}\right)\sin\theta\cos\theta  
\quad \hbox{and} \quad 
q=\frac14\frac{\Omega^2 R^3}{M J_2}
\end{equation}
are the principal forms of the its Ricci tensor.

To construct the Principal transform of the Fermat quo-tensor as we
defined it in subsection 2.2 we need to find the
three scalars $c_a$ such that the metric:

\begin{equation}
\label {4.7}
d\bar s^2=c_1^2(\hat\theta^1)^2+c_2^2(\hat\theta^2)^2+c_3^2(\hat\theta^3)^2
\end{equation}
satisfies the two sets of equations \ref{2.25}, with $\kappa=0$, and
\ref{2.27}. These coefficients have been found to be at the desired
approximation:

\begin{eqnarray}
\label {4.8}
c_1=1-\frac{M}{r}\left(1+\frac15\frac{R_1^2}{r^2}\right)+
\frac{M J_2
R^2}{r^3}\left(6+\frac{6}{5}\frac{R^2}{r^2}-\left(5+\frac{9}{5}\frac{R^2}{r^
2}
\right)\sin^2\theta\right) \nonumber \\
+\Omega^2 r^2\left(-\frac{27}{5}+\frac{3}{10}\frac{R^2}{r^2}
+\left(\frac{15}{2}-\frac{3}{10}\frac{R^2}{r^2}\right)\sin^2 \theta\right)
\end{eqnarray}

\begin{eqnarray}
\label {4.9}
c_2=1-\frac{M}{2r}\left(3-\frac{1}{5}\frac{R_2^2}{r^2}\right)+
\frac{M J_2 R^2}{r^3}\left(1-\frac{3}{5}\frac{R^2}{r^2}-\left(\frac{19}{4}
-\frac{21}{20}\frac{R^2}{r^2}\right)\sin^2\theta\right) \nonumber \\
+\Omega^2 r^2\left(-\frac{19}{5}
+\left(\frac{13}{2}+\frac{3}{10}\frac{R^2}{r^2}\right)\sin^2 \theta\right)
\end{eqnarray}

\begin{eqnarray}
\label {4.10}
c_3=1-\frac{M}{2r}\left(3-\frac{1}{5}\frac{R_2^2}{r^2}\right)+
\frac{M J_2 R^2}{r^3}\left(1-\frac{3}{5}\frac{R^2}{r^2}-\left(\frac{9}{4}
+\frac{3}{4}\frac{R^2}{r^2}\right)\sin^2\theta\right) \nonumber \\
+\Omega^2 r^2\left(-\frac{19}{5}+4 \sin^2 \theta\right)
\end{eqnarray}
where $R_1$ and $R_2$ are in this approach two free parameters of
the order of the mean radius $R$ of the Earth. The values of these 
coefficients provide the basis for a fresh new discussion about the
anisotropy of space in the neighborhood of the surface of the Earth
and the means to measuring it.

From our general principle of interpretation of subsection 2.2
according to which the coefficients $c_a$ are in particular the
velocities of light along the principal directions \ref{4.5} and from
\ref{2.63}, it follows at the required approximation, that the velocity of
light in the direction with components $\gamma_a$ with respect to the
principal triad \ref{4.5} is:

\begin{equation}
\label {4.13}
v=1+\gamma^2_1\alpha_1+\gamma^2_2\alpha_2+\gamma^2_3\alpha_3 \quad
\hbox{with} \quad \alpha_a=1-c_a  
\end{equation}

Since the interferometer in the Michelson-Morley experiment is kept
horizontal the relevant expression in this case is:

\begin{equation}
\label {4.14}
v=1+\alpha_2\cos^2A+\alpha_3\sin^2A=
1+\frac12(\alpha_2+\alpha_3)+\frac12(\alpha_2-\alpha_3)\cos 2A 
\end{equation}
where A is the azimuth of the direction $\gamma_a$, and where in
particular the quantity which measures the 
anisotropy is:

\begin{equation}
\label {4.11}
a_2=\frac12(c_2-c_3)=-\frac{1}{10}\left(\frac{11 M J_2}{R}-14 \Omega^2
R^2\right)\sin^2\theta
\end{equation}
on the surface of the Earth at the colatitude $\theta$. Taking into
account the following values: $M=0.00444 \hbox{\, m}$, $R=6378164
\hbox{ m}$, $\Omega=2.434\times 10^{-13} \hbox{\, m}^{-1}$, and
$J_2=0.0010826$ we have: 

\begin{equation}
\label {4.12}
a_2=2.5\times 10^{-12}\sin^2 \theta
\end{equation}
This dependence on the colatitude is a signature that it will help in
testing this effect.

Up to know only one experiment, \cite{Brillet-Hall}, has been performed
with a sufficient sensitivity to test this result. In this experiment
the coefficients $a_2$ and $b_2$ in the following expression:

\begin{equation}
\label {4.15}
v=\frac12 a_0+a_2\cos 2A+b_2\sin 2A
\end{equation} 
are measured, but since the paper does not indicate what was the
experimental origin of the angle $A$ only the value of
$\sqrt(a_2^2+b_2^2)$ can be considered as meaningful. The result
which is mentioned is $2.1\times 10^{-13}$ at a colatitude of
$50^\circ$. The corresponding result
derived from \ref{4.12} is $1.5\times 10^{-12}$. The agreement here is
only qualitative. A better one is obtained with a slightly different
approach (See the footnote at the beginning of this section and the
references therein). In our opinion it is an urgent task that
somebody repeats this experiment with the required sensitivity.

\section{The definition of the quantum vacuum in Milne's universe}

The two preceding applications made an explicit use of the
interpretation we proposed in section 2 embodied in the concept of
meta-rigid motions. They made use also of principal transformations
of the Fermat tensor into a constant curvature metric, thus
implementing the property of free mobility of idealized rigid bodies.
Neither application necessitated the use of any particular
synchronization, contrary to the application of this section and the
two following ones.

We consider here the problem of defining the quantum vacuum of a
quantized scalar field in Cosmology. But instead of keeping the
subject rather general, as it was done in \cite{Vacuum}, we describe
here the problem in the context of Milne's universe, because its
geometry, which is locally flat, raises some other problems related to
the main subject of\ this contribution,\footnote{Another particular
example can be seen in \cite{Mallorca}}.

Let us consider Minkowski's space-time $\cal M$ referred to a galilean
frame
of reference and cartesian coordinates $x^\alpha$. Let $E$ be an
event on it and let us consider the interior of the future-pointing
light-cone with vertex $E$. This defines the manifold $\cal U$ of Milne's
universe. Its metric being that of the flat space-time corresponding
to its immersion on $\cal M$. Let us consider on $\cal U$ the
congruence $\cal R$ of time-like geodesics passing through $E$. An
appropriate
parameterization of $\cal R$ is:

\begin{equation}
\label {5.1}
x^0=t, \quad x^i=z^i Ht
\end{equation}  
where $H$ is an arbitrary constant with dimensions $T^{-1}$. Using
$(t,z^i)$
as adapted coordinates the line element of flat space-times becomes:

\begin{equation}
\label {5.2}
ds^2=-(1-H^2 r^2)dt^2 + H^2\delta_{ij}(2tz^idt + t^2dz^i)dz^j
\end{equation}
where $r^2=\sum (z^i)^2$. The following time-gauge transformation:

\begin{equation}
\label {5.3}
t\rightarrow t(1-H^2r^2)^{1/2}
\end{equation}
and the use of polar coordinates instead of cartesian ones brings the line 
element \ref{5.2} to the following form:

\begin{equation}
\label {5.4}
ds^2=-dt^2+d\hat s^2, \quad 
d\hat s^2= H^2t^2d\bar s^2
\end{equation}
with:

\begin{equation}
\label {5.16}
d\bar s^2=\frac{1}{1-H^2r^2}\left(\frac{dr^2}{1-H^2r^2}
+r^2(d\theta^2+\sin^2d\varphi^2)\right)
\end{equation}
which after the final transformation of the coordinate $r$:

\begin{equation}
\label {5.5}
r \rightarrow r(1-H^2r^2)^{-1/2}
\end{equation}
becomes:

\begin{equation}
\label{5.5x}
d\bar s^2=\frac{dr^2}{1-H^2r^2}+r^2(d\theta^2+\sin^2d\varphi^2)
\end{equation}
We recognize in \ref{5.4} with $d\bar s^2$ given by \ref{5.5x} one of
the familiar forms of a Robertson-Walker metric. In this case, i.e.
the Milne's universe, the scale factor is $Ht$ and the space
curvature is $-H^2$.  

Since the Laplacian $\hat \triangle$ defined in \ref{2.14} coincides
in this case with $\bar \triangle$, the Laplacian corresponding to
the time independent principal transform metric \ref{5.5x}, it follows
that the congruence $\cal R$ is quo-harmonic. Although quo-harmonic
coordinates are not in this case the more convenient coordinates to
use, even in its polar form.

Notice also that since the family of hyper-surfaces $t=const$ is
orthogonal to $\cal R$ the synchronization that it defines is a chorodesic
one. It is not a geodesic one because Born's deformation rate of $\cal R$
is not zero. The time scale is the corresponding {\it TAC} as we defined it
in section 2. In this case it does not depend on the world-line of
$\cal R$ to which it is associated.

It is not important here to decide whether or not Milne's universe is
a good model of our real universe. It is sufficient to recognize that
a real universe could have existed for which Milne's universe would have
been an acceptable model. 
Two remarks are relevant here: i) the first one is that although the
local geometry of both the Milne's and Minkowski's universe is the
same their physical meaning is quite different. And anybody, who would
say that they describe the same physical reality on the basis that
the coordinate singularity at $r=0$ could be avoided by extending the
geodesically incomplete Milne's universe to span the full Minkowski's
universe, would mathematically be correct but absolutely wrong from a
physical point of view. ii) the second remark is that even from the local
point of view Milne's and Minkowski's universes describe different
physical realities if the theory of frames of reference that we
presented in section 2 is accepted. In fact one has to consider
that the congruence $\cal R$ which we considered to build Milne's model is
a
perfect meta-rigid motion in this universe while it is not in
Minkowski's one.

The Klein-Gordon equation in Minkowski's space-time referred to a
galilean frame of reference is :

\begin{equation}
\label {5.16}
(-\partial_t^2+\triangle
-\frac{m^2}{\hbar^2})\psi=0
\end{equation}
where $\hbar$ is the reduced Plank's constant, $m$ is the mass of
the quantum of the field and $\triangle$ is the Laplacian
corresponding to the euclidean metric.

The positive(negative) energy modes are the
solutions of \ref{5.6} which have the following form:

\begin{equation}
\label {5.7}
\varphi_\epsilon(x^\alpha, \vec k)=(2\pi\hbar)^{-3/2}
u_\epsilon(t,\vec k)\exp({\frac{i}{\hbar}\vec k \vec x}), 
\end{equation}
with:

\begin{equation}
\label {5.8}
u_\epsilon(t, \vec k)= 
(2\omega)^{-1/2}\exp({-\frac{i}{\hbar}\epsilon \omega t}), \quad 
\omega(\vec k)=+({\vec k}^2+m^2)^{1/2}
\end{equation}
where $\epsilon=1\,(-1)$. This decomposition of the space of
solutions in two distinct sub-spaces is complete, intrinsic and it is
what is needed to introduce the concept of particles and
anti-particles as quanta of a quantized scalar field. It is usually
referred to as its quantum vacuum.

In Milne's universe the Klein-Gordon equation for a classical field
$\Psi$ is:

\begin{equation}
\label {5.6}
(-\partial_t^2-3\dot{\sigma}\partial_t+e^{-2\sigma}\bar\triangle
-\frac{m^2}{\hbar^2})\psi=0
\end{equation}
where $\sigma=\ln (Ht)$ and $\bar\triangle$ is the laplacian of the
constant curvature metric $d\bar s^2$. 
A complete set of solutions of \ref{5.6} can be
obtained in two steps: i) separating first the time dependence and the
space dependence, i.e. by looking for solutions of the following
form:

\begin{equation}
\label {5.9}
\Psi(x^\alpha)=u(t)\psi(x^i, \vec k)
\end{equation}
$\psi(x^i, k^2)$ being a complete set of eigen functions of the
Laplace operator $\bar\triangle$ of the metric \ref{5.5x}:

\begin{equation}
\label {5.10}
\bar\triangle  \psi(x^i, \vec k)=-(k^2/\hbar^2)\psi(x^i, \vec k)
\end{equation}
and ii) splitting the space of solutions $u(t, k^2)$ of the following
equation:

\begin{equation}
\label {5.11}
\hbar^2\ddot u+3\hbar^2t^{-1}\dot u + \omega^2 u=0, \quad 
\omega^2=\frac{k^2}{H^2t^2}+m^2
\end{equation}
in two groups. To do that 
we proposed in \cite{Vacuum} to solve the problem in two steps: a) to
solve first the Riccati equation:

\begin{equation}
\label {5.12}
i\hbar \dot f+f^2+3 i \hbar t^{-1} f-\omega^2=0
\end{equation}
and b) to solve after that the equation:

\begin{equation}
\label {5.17}
i\hbar \dot u=fu 
\end{equation}
$f$ being a solution of Eq. \ref{5.12}. This guarantees that the first
order equation \ref{5.17} is an order reduction of \ref{5.11} in the
sense that every solution of \ref{5.17} is a solution of this equation
but not vice versa. The positive energy solutions of \ref{5.11} are
then by definition the solutions of \ref{5.17} with $f$ being the solution
of \ref{5.12} satisfying the following limit condition with $\epsilon=+1$ 

\begin{equation}
\label {5.13}
\lim_{\hbar \rightarrow 0}f_\epsilon=\epsilon \omega.
\end{equation}
The negative energy solutions are those corresponding to
$\epsilon=-1$

If $m=0$ the solutions are very easy to obtain. They are

\begin{equation}
\label {5.14}
f_\epsilon=bt^{-1}\quad \hbox{with} \quad 
b=-i\epsilon\hbar+(-\hbar^2+k^2/H^2)^{1/2}
\end{equation}
And we have thus:

\begin{equation}
\label {5.15}
u_\epsilon(t,\vec k)=A(t_0,\vec k)\exp(-\frac{ib}{\hbar}\ln(t/t_0))
\end{equation}
where $t_0$ is an arbitrary constant and A is a function of $t_0$ and
$k^2$ to be fixed with a normalization condition.

As we see the quantum vacuum of Minkowski's universe referred to a
galilean frame of reference is quite different from that of Milne's
universe referred to the frame of reference $\cal R$ defined above.
Nevertheless it makes sense to compare \ref{5.15} to \ref{5.8} (with
$m=0$) because in either case the
frame of reference which is used correspond to the same definition:
The meta-rigid motion is a quo-harmonic congruence, the
synchronization is a chorodesic one, and the time scales are
similarly defined.

\section{The missing group}

In 1872 Felix Klein gave a celebrated definition saying that a
geometry is a theory of the invariants under a particular group of
transformations. The geometry of both Classical physics and Special
relativity will fit this definition because the laws of physics in
both cases are invariant under the group of Galileo or under the
Poincar\'e group respectively. But,  which is the group of transformations
defining the geometry of space-time as we use it in general
relativity ?. A possible honest answer is to to think about the
isometry group of the space-time metric and say that in general there is
none, or that when there is one it is usually very small and does
not  fit very well in Klein's definition. Does this mean that there
is a missing group? There is of course also the possibility of 
saying that there is no missing group and that 
one has to think
of the group of diffeomorphisms of the space-time manifold
 as a group of symmetry, which in our opinion it is not.  I would
like to insist here on a suggestion I made a few years ago in
\cite{Salas} by saying that for any space-time we should find a group
more general than the group of isometries and more meaningful than 
the group of diffeomorphisms. 

At the end of this section we describe a non trivial example illustrating
this
idea. The first part of the idea consists in realizing that the more
fundamental group of classical mechanics is not the Galileo group but
a much larger non-Lie group, namely the group of rigid motions which
transform the cartesian and time coordinates of two rigid frames of
reference one into another:

\begin{equation}
\label {6.1}
t'=t-A^0, \qquad x'^i= R^i_j(t)\left[x^j- A^j(t)\right],
\end{equation}
where $R^i_j(t)$ is a rotation matrix. This group contains as sub-groups
the
euclidean group in three dimensions:

\begin{equation}
\label {6.2}
x'^i= R^i_j(x^j-S^j), \qquad R^i_j, S^j: Constants
\end{equation}
and the inhomogeneous Galileo group:

\begin{equation}
\label {6.3}
 t'=t-A^0, \quad x'^i= R^i_j\left(x^j-V^jt-A^j\right), 
\qquad R^i_j, V^j, A^\alpha: Constants
\end{equation}
which are the sub-groups people usually think of when considering the
basic symmetries in classical physics.

Let us consider the class $\cal N$ of space-times for which there exists a
system of coordinates such that their line-element can be written as:

\begin{equation}
\label {6.4}
ds^2=-(1-2U)dt^2+2U_idt dx^i+\delta_{ij}dx^idx^j
\end{equation}
where $U$ and $U_i$ are functions of $t$ and $x^i$. The group of
rigid motions acts on a line-element of this sort as a group of
generalized isometries in the sense that it maps the class $\cal N$
into itself, keeping invariant the restriction of the line-element on
any hyper-surface $t=const.$. Our quest for the missing group stars
with this remark which led also in \cite{Bel90} to a generalized
Newtonian theory of gravitation in classical mechanics for which the
meaningful covariance is the generalized invariance under the group
of rigid motions.

Let us consider now a given space-time $M$ and let us assume that we
know the class $\cal C$ of quo-harmonic congruences satisfying every
necessary condition to qualify as meta-rigid motions of $M$. This
class may contain some, or all, of the isometries of $M$; some, or
all, of Born's rigid congruences. We wrote: some or all, because
global conditions or physical requirements may disqualify some
quo-harmonic congruences as meta-rigid motions. In general $\cal C$
will contain also an infinite number of other congruences, possibly
labeled by arbitrary functions and parameters. Let us assume also
that for each of these congruences the principal transformation of
the Fermat tensor is un-ambiguously defined by every necessary global
condition and therefore there exists a well defined system of
quo-harmonic coordinates $z^i$ corresponding to the constant
curvature image of the Fermat tensor.  Let us assume finally that a
chorodesic synchronization has been chosen for each congruence of
$\cal C$ and therefore the time coordinate $z^0=\tau$ is sufficiently
well defined, up to a time sub-group transformation. If $\cal R$ and
${\cal R}^\prime$ are two meta-rigid motions of $\cal C$ labeled by
$L$ and $L^\prime$ we may consider the coordinates
$(\tau^\prime,z^{\prime i})$ of an event with respect to ${\cal
R}^\prime$ as functions of the coordinates $(\tau,z^i)$ of the same
event with respect to $\cal R$. We write this coordinate
transformation symbolically:

\begin{equation}
\label {6.5}
z^{\prime \alpha}=\varphi^\alpha_L(z^\beta, L^\prime)
\end{equation}
By definition this family of transformations satisfy a composition
law:

\begin{equation}
\label {6.9}
\varphi^\alpha_{L^\prime}(\varphi^\beta_L(z^\gamma,L^\prime),L^{\prime 
\prime})= \varphi^\alpha_L(z^\beta,
\lambda(L,L^\prime,L^{\prime\prime}))
\end{equation}
where $\lambda(L,L^\prime,L^{\prime\prime})$ is some composition law
in the space of labels.
  
We analyze below with more detail the family of transformations
\ref{6.5} in
a particular case. Let $M$ be Minkowski's space-time with line element
in a system of galilean coordinates $x^\alpha$:

\begin{equation}
\label {6.10}
ds^2=-dt^2+\delta _{ij}dx^idx^j.
\end{equation}
Let $W$ be
a time-like world-line with parametric equations, and unit time-like
tangent vector:

\begin{equation}
\label {6.6}
x^\alpha=y^\alpha(\tau), \quad w^\alpha(\tau)=\dot y^\alpha(\tau)
\end{equation}
where $\tau$ is proper-time along $W$ measured from a fixed 
origin on $W$ with galilean coordinates $y^\alpha_0$. 

There is one and only one irrotational Born congruence
containing $W$ and this is the Fermi-congruence with base-line $W$
and parametric equations:

\begin{equation}
\label {6.11}
x^\alpha = e_i^\alpha (\tau )z^i+ y^\alpha
(\tau ),
\end{equation}
where $e_i^\alpha (\tau )$ is a triad of orthonormal vector
fields,
orthogonal to $w^\alpha(\tau)$, Fermi propagated along $W$, i.e.
satisfying the differential equations:

\begin{equation}
\label {6.12}
\dot e_i^\alpha =-w^\alpha b_\rho e_i^\rho \quad
b_\rho=\dot w_\rho.
\end{equation} 

Each congruence of $\cal B$ can be labeled by three functions of one
argument: $y^i(\tau)$ and one parameter: $y^0(0)$. Since if we know 
$y^i(\tau)$ we know also $u^i(\tau)={\dot u}^i$, and since $u^\alpha$
has to be unitary this means then that we know also $u^0$.
Integrating the differential equation ${\dot y}^0=u^0$ with the
initial condition $y^0(0)$ yields then the function $y^0(\tau)$ and
the parametric equations of the base-line of the congruence are
known. This proves that the family of congruences $\cal B$ has the
same number of degrees of freedom: three arbitrary functions
$y^i(\tau)$ of one variable and a parameter $y^0(0)$, as the family of
irrotational rigid motions of classical physics. 

Differentiating Eqs. \ref{6.11} and substituting the result in
\ref{6.10} the line
element of $M$ becomes \cite{Moller}:

\begin{equation}
\label {6.7} 
ds^2=-[1+a_k(\tau)z^k]^2d\tau^2+\delta_{ij}dz^idz^j, \quad
a_k(\tau)=e^\alpha_k(\tau)b_\alpha(\tau).  
\end{equation}
$a_k(\tau)$ is the curvature, i.e. the intrinsic acceleration, of $W$
which parametric equations in the adapted coordinates are $z^i=0$.

The form of the line-element \ref{6.7} proves that the Fermat tensor
is euclidean, meaning that the principal transformation in this case
is trivial; it proves moreover that the Fermi coordinates of space
are the corresponding quo-harmonic, in this case cartesian,
coordinates. On the other hand the $\tau=const.$ hyper-surfaces
define a chorodesic, in this case geodesic, synchronization.

In \cite{Salas} we defined a compound description of a metric
$g_{\alpha\beta}(z^\rho)$ as a description of this metric by 10
functions $\bar g_{\alpha\beta}(x^\rho,f_{(L)})$ where $f_{(L)}$ is a
set of functions of $x^\rho$ belonging to a restricted functional
space $\cal F$. The metric \ref{6.7} is an example of compound
description of the Minkowski metric where the functional space
$\cal F$ is the set of the triads $a_k(\tau)$ of functions of one
variable. 

We shall say that the class of congruences $\cal C$
is isometrically complete if a compound description of the space-time
metric exist such that we have:

\begin{equation}
\label {6.13}
g_{\alpha\beta}(x^\rho; L)=\bar g_{\alpha\beta}(x^\rho;
f_{L}(x^\sigma))
\end{equation}
where for each $L$, i.e. for each quo-harmonic congruence of $\cal
C$, $x^\rho$ is an admissible system of quo-harmonic 
coordinates\,\footnote{We call also quo-harmonic a time coordinate 
of the {\it TAC} scale corresponding to the frame of reference 
being considered} and $f_{(L)}(x^\sigma)$ is an appropriate set of $\cal
F$.
By construction  the class $\cal B$ of irrotational Born rigid
motions in Minkowski's space-time \ref{6.11} is isometrically complete
and the corresponding compound description of the space-time metric
is \ref{6.7}

The introduction of the concept of compound description of a
space-time metric allows to interpret the coordinate transformations
\ref{6.5} as a group of transformations defining what we called in
\cite{Salas} {\it Generalized isometries}. In fact, let us consider
the family of coordinate transformations that leave invariant the
compound description of an isometrically complete class of
quo-harmonic congruences $\cal B$. This family of transformations,
which coincides with the family of transformations \ref{6.5}, is
a group, actually an infinite dimensional one in general. It must be
understood though that the space on which these
transformations act as a group is the product $M\times \cal F$, or some
open
neighborhood of it. Underlining this it may be convenient to use the
notation:

\begin{equation}
\label {6.14}
z^{\prime\alpha}=\psi^\alpha(z^\beta, f_{(L)}; \Lambda) \quad
f_{(L^\prime)}=\chi_{(L^\prime)}(f_{(L^\prime)}, \Lambda)
\end{equation}
where $\Lambda$ designates a functional space of parameters.

In \cite{Salas} we defined the {\it Born's group} as being the group above
for the class of
Born congruences $\cal B$ in Minkowski's space-time. In other words
the Born's group is the group of transformations that leave invariant
the compound description \ref{6.7}. 

For each set of functions
$a_k(\tau)$ the family of transformations that leave invariant the
compound description and the functions $a_k(\tau)$ themselves, defines a
realization of the Poincar\'e group. 

A particular sub-group of the Born group deserves a particular
attention. It is the {\it Time sub-group} which leaves invariant a
particular congruence of $\cal B$ without leaving invariant the set
of functions $a_k(\tau)$. It is the group of transformations of Fermi
coordinates that switch from a world-line seed of the congruence to
another. The analytic realization of this finite dimensional sub-group is:

\begin{equation}
\label {6.15}
z^{^{\prime }i}=z^i-\lambda ^i, \quad 
\tau ^{^{\prime }}=\tau +v_k(\tau)\lambda ^k, \quad
a_k^{^{\prime }}(u,\lambda ^i)={\frac{a_k(u)}{{1+a_j(u)\lambda ^j}}} 
\end{equation}  
where:

\begin{equation}
\label {6.16}
v_k(\tau )=\int_0^\tau a_k(u)du 
\end{equation}
  
\section{The froth of the universe}

Let us consider a general Robertson-Walker metric:

\begin{equation}
\label {7.1}
ds^2=-c^2dt^2+F^2(t)d\bar s^2, \quad d\bar s^2=N^2(r)\delta_{ij}dx^idx^j, 
\end{equation} 
where:

\begin{equation}
\label {7.2}
N(r)=\frac{1}{1+kr^2/4}, \quad r=\sqrt{\delta_{kl}x^kx^l}.
\end{equation}
We shall assume that the scale factor $F$ is dimensionless, that $t$
has dimensions of time $T$ and that $r$ has dimensions of space $L$. 
The curvature constant $k$ will have therefore dimensions $L^{-2}$
and the universal constant $c$\,\footnote{In this section we do not
use normalized units} dimensions $LT^{-1}$.

The line-element \ref{7.1} can be written also in the following form:

\begin{equation}
\label {7.3}
d\tau^2=dt^2-\frac{1}{c_{eff}^2}d\bar s^2 \quad \hbox{with} \quad
c_{eff}=c/F(t)
\end{equation}
which suggests that any local process, both in time and in space, will
evolve as if the Universe were flat with an effective speed of light
equal to $c_{eff}$. It suggests also the idea of giving
different names to the universal constant $c$ and to $c_{eff}$.
Several years ago, \cite{Bel90}, we proposed to call R\"omer's
constant the universal constant $c$ which allows the
conversion of space to time and vice versa. Instead $c_{eff}$ will be
here the speed of those processes mediated by zero mass
particles, in particular the propagation of light.
 
Let us consider the congruence $\cal R$ with world-lines $x^i=const.$. The
Fermat metric of this congruence is:

\begin{equation}
\label {7.4}
d\hat s^2=(1/c_{eff})^2d\bar s^2
\end{equation}
By the same arguments we gave in section 5 it follows from \ref{7.3}
that $\cal R$ is a quo-harmonic congruence, that the family of
hyper-surfaces $t=constant$ define a chorodesic synchronization, and 
that the coordinate $t$ is a {\it TAC} coordinate for every world-line
of $\cal R$.

According to the definition we gave in section 2 the principal 
transformation of \ref{7.4} is the metric $d\bar s^2$ defined above
and therefore the three coefficients $c_a$ are all equal to
$c_{eff}$. The local heuristic interpretation of this function given above
is therefore substantiated by the theory of principal transformations,
becoming a global and theoretically justified interpretation.
 
If we assume to know the scale Factor $F(t)$ and the remaining
parameters of the model then Einstein's field equations:

\begin{equation}
\label {7.5}
S_{\alpha\beta}-\Lambda g_{\alpha\beta}=\frac{8\pi G}{c^2}T_{\alpha\beta}
\end{equation}
will allow to calculate $T_{\alpha\beta}$. The energy-momentum tensor
thus defined is however a rather polymorphic object that can be
written as if we were dealing with a perfect fluid:

\begin{equation}
\label {7.6}
T_{\alpha\beta}=\rho u_\alpha u_\beta + p\hat g_{\alpha\beta}, \quad
\hat g_{\alpha\beta}=g_{\alpha\beta}+ u_\alpha u_\beta
\end{equation}  
where the pressure, $p$, and the mass density$, \rho$, are functionals of
$F(t)$, but it can be written also as if we were dealing with a
scalar field source:

\begin{equation}
\label {7.7}
T_{\alpha\beta}=\partial_\alpha\Phi\partial_\beta\Phi
-g_{\alpha\beta}[\partial_\mu\partial^\mu\Phi-V(\Phi)]
\end{equation}
with $\Phi$ and $V(\Phi)$ being some other functionals of $F$. It is
also possible, among many other possibilities, to think of
$T_{\alpha\beta}$
as being a superimposition of expressions of the type \ref{7.6} and
\ref{7.7}.

Of course a cosmologist works the other way around. He wants to
derive $F$ from a single polymorphic instance of $T_{\alpha\beta}$ or
from a few ones, depending on the stage of the evolution of the
universe he considers or the degree of `realism' he wants to include
in his model. The so called Standard model assumes that the history
of the universe can be divided in two epochs. It assumes, as the
Bible, that at the beginning God said: {\it Fiat lux}, and the
universe became filled with radiation, or equivalently that
$T_{\alpha\beta}$ took the form \ref{7.6} with the equation of state  
$p=1/3\rho c^2$. It assumes also that during the, still lasting,
second epoch the Universe became matter dominated which means that
$T_{\alpha\beta}$ still has the form \ref{7.6} but this time with
$p=0$.  The so called Inflationary models are a variant of the
Standard model where one assumes that at some epoch the
Universe was dominated by some scalar field and therefore
$T_{\alpha\beta}$ could be interpreted as in  \ref{7.7}. All the details
about this scalar field and the mechanisms of transition from one
epoch to another rely heavily on our knowledge of many other branches
of physics. This makes these models rather `realistic' but at the same
time they are probably too cumbersome and unnecessarily complicated
for what observational cosmology has to offer today.      

In \cite{Froth} we presented a model which is conceptually and
technically very simple and seems to offer many of the properties of
more elaborate models. The model is based on what we said before at 
the beginning of this section about the meaning of $c_{eff}$. The
idea consists in observing that the equation of state $p=1/3\rho c^2$
is the equation of state of a gas of zero mass particles, and that this
equation is derived in the frame-work of special relativistic kinetic
theory, assuming that the box containing the gas is at rest with
respect a galilean frame of reference. In which case
the speed of light in vacuum coincides with the
universal R\"omer's constant. The distinction we made above
between these two concepts, the heuristic argument we mentioned, and
a more detailed analysis included in \cite{Froth} suggest that the
equation of state of such a gas should be instead:

\begin{equation}
\label {7.8}
p=\frac{1}{3}\rho c^2/F^2=1/3\rho c_{eff}^2
\end{equation} 

Solving Einstein's equations with this equation of state gives for the
density:

\begin{equation}
\label {7.9}
\rho=\rho_0 F^{-3}\exp[1/2(1/F^2-1)]
\end{equation}
where $\rho_0$ is the present density if we assume that the present
value of $F$ is $1$, i.e. we assume that the value of $c$ is equal to
the value $c_{eff}$ now. Any other choice would just shift the
present value of $F$.

It can be seen that for very small values of $F$, $T_{\alpha\beta}$
viewed as  an energy-momentum
tensor having the form \ref{7.7} corresponds to a scalar field:

\begin{equation}
\label {7.10}
\Phi=\frac{1}{F}
\end{equation}     
with potential:

\begin{equation}
\label {7.11}
V(\Phi)=-\frac{\rho_0}{6}\Phi^5\exp[(1/2)(\Phi^2-1)]
\end{equation}
On the other hand, assuming that $F$ can reach sufficiently large values,
the model behaves as if $T_{\alpha\beta}$ had the form \ref{7.6} of a
perfect fluid with:

\begin{equation}
\label {7.12}
p=0
\end{equation} 
i.e. as the energy-momentum of dust matter.

Although it is possible to conceive a model of the Universe for which
the equation of state \ref{7.8}would hold during a small fraction of its
history after the big-bang and recur to more common ideas afterwards,
it is tempting to consider a model for which the equation of state
would hold from the beginning to the end. This model depends then on
two free parameters, the curvature constant $k$ and the cosmological
constant $\Lambda$. On the other hand to determine completely the
model we need to know the present value of $\rho_0$. This is a
crucial point, because the value to be assigned to this quantity depends
on the interpretation that one wishes to give to the fluid described
by the equation of state \ref{7.8} at present time. Two possibilities can
be considered: i) We can assume that this fluid is 
equivalent to the matter content of the universe, in which case
the value of $\rho_0$ should be
estimated as it is usually done: counting galaxies, guessing masses
or else.
This would be a complicated model not very different from
standard models and we do not consider it here anymore. ii) We can
assume that ordinary matter is hierarchically organized with a
fractal dimension less than 3 which means that it would be weightless
compared to the energy of the background blackbody radiation at
temperature $2.7\,K$. In other words matter would be just {\it The
froth of the Universe}. This is a very simple and viable model to
which Stefan's law assigns the following value to the present
density:

\begin{equation}
\label {7.13}
\rho_0=4.5\times 10^{-31}\, \mbox{kg/m}^3
\end{equation}  
As usual two other quantities have to be derived from observation.
Namely the Hubble constant $H_0$ and the deceleration parameter $q_0$.
Considering as an example the following values:

\begin{equation}
\label {7.14}
H_0=75\, \mbox{km/s/Mpc}, \quad q_0=0.1
\end{equation}        
the model yields the following values for $\Lambda$ and $k$:

\begin{equation}
\label {7.15}
\Lambda=-2.\times 10^{-51} \mbox{m}^{-2}, \quad 
k=-7.3\times 10^{-51} \mbox{m}^{-2}
\end{equation}

\section*{Appendix: Some definitions and notations}
Given any space-time with line element:

\begin{equation}
\label {2.1}
ds^2=g_{\alpha\beta}(x^\rho)dx^\alpha dx^\beta \quad 
\alpha,\beta,\cdots=0,1,2,3 \quad x^0=t,
\end{equation}
let us consider a time-like congruence $\cal R$, $u^\alpha$ being its unit
tangent vector field ($u_\alpha u^\alpha=-1$). The {\it Projector} into the
plane orthogonal to $u^\alpha$ is:

\begin{equation} 
\label {8.14}
\hat g_{\alpha \beta }=g_{\alpha \beta }+u_\alpha u_\beta  
\end{equation}

By definition the {\it Newtonian field} is the opposite to the acceleration
field:

\begin{equation}
\label {8.10}
\Lambda_\alpha=-u^\rho \nabla _\rho u_\alpha, 
\quad \Lambda_\alpha u^\alpha=0.
\end{equation} 
The {\it Coriolis field}, or the rotation rate field, is the skew-symmetric
2-rank tensor orthogonal to $u^\alpha $:

\begin{equation}
\label {8.11}
\Omega _{\alpha \beta }=\hat \nabla _\alpha u_\beta -\hat
\nabla _\beta u_\alpha ,\quad \Omega _{\alpha \beta }u^\alpha =0. 
\end{equation}
where: 

\begin{equation}
\label {8.11bis}
\hat \nabla _\alpha u_\beta \equiv \hat g_\alpha ^\rho \hat g_\beta ^\sigma
\nabla _\rho u_\sigma. 
\end{equation}
And {\it Born's deformation rate field} is the symmetric 2-rank
tensor orthogonal to $u^\alpha $:

\begin{equation}
\label {8.12}
\Sigma _{\alpha \beta }=\hat \nabla _\alpha u_\beta +\hat
\nabla _\beta u_\alpha ,\quad \Sigma _{\alpha \beta }u^\alpha =0, 
\end{equation}

Let $x^\alpha$ be a system of adapted coordinates,
i.e., such that $u^i=0$. We use the following
notations:

\begin{equation}
\label {2.2}
\xi = \sqrt{-g_{00}}, \qquad  \varphi_i = \xi^{-2}g_{0i} ,
\end{equation}
and:

\begin{equation}
\label {2.3}
\hat g_{ij} = g_{ij} +
\xi^2\varphi_i\varphi_j \quad .
\end{equation}
which we shall call the {\it Fermat} quo-tensor of the congruence.
Here and below quo-tensor refers to an object, well defined on the
quotient manifold ${\cal V}_3={\cal V}_4/\cal R$, 
which covariant components are
the space components of a tensor of ${\cal V}_4$ orthogonal to
$u^\alpha$. 

The {\it Newtonian field} is then the quo-vector:

\begin{equation}
\label {2.4}
\Lambda_i = -(\hat\partial_i \ln\xi +\partial_t \varphi_i), \quad
\hat\partial_i\cdot \equiv
\partial_i\cdot + \varphi_i\partial_t\cdot
\end{equation}
The {\it Coriolis field}, or Rotation rate field, is the skew-symmetric
quo-tensor:

\begin{equation}
\label {2.7}
\Omega_{ij}=\xi(\hat\partial_i \varphi_j-\hat\partial_j \varphi_i)
\end{equation}
and the {\it Born's deformation rate field} is the symmetric quo-tensor:

\begin{equation}
\label {2.5}
\Sigma_{ij}=\hat\partial_t\hat g_{ij}, 
\quad \hat\partial_t=\xi^{-1}\partial_t
\end{equation}

To these familiar geometrical objects it is necessary 
to add the following ones, \cite{Zelmanov},
\cite{Cattaneo}, \cite{Ida}:

\begin{equation}
\label {2.8}
\hat\Gamma^i_{jk} = \frac12 \hat g^{is}(\tilde\partial_j\hat g_{ks}
+ \tilde\partial_k\hat g_{js} -\tilde\partial_s\hat g_{jk}) ,
\end{equation}
which are the {\it Zel'manov-Cattaneo symbols}, and:

\begin{equation}
\label {2.9}
\hat R^i_{jkl} = \tilde\partial_k\hat\Gamma^i_{jl} -
\tilde\partial_l\hat\Gamma^i_{jk} + \hat\Gamma^i_{sk}\hat\Gamma^s_{jl} -
\hat\Gamma^i_{sl} \hat\Gamma^s_{jk}
\end{equation} 	
which is the {\it Zel'manov-Cattaneo quo-tensor},

\end{document}